%% file: h2qed.tex
\documentclass[aps,pra,amsmath,amssymb,amsfonts,reprint,twocolumn]{revtex4-2}
\usepackage[T1]{fontenc}
\usepackage{times}
\usepackage{graphicx}
\usepackage{scrextend}
\usepackage{dcolumn}
\usepackage{bm}
\usepackage{bbold}
\usepackage{dcolumn}
\newcolumntype{d}[1]{D{.}{.}{#1}}
\newcommand\Tstrut{\rule{0pt}{2.6ex}} 
\newcommand\Bstrut{\rule[-0.9ex]{0pt}{0pt}}
\allowdisplaybreaks

\usepackage[table]{xcolor}

\begin{document}
\preprint{Version 1.0}

\title{Leading-order QED effects in the ground electronic state of molecular hydrogen}

\author{Micha{\l} Si{\l}kowski}
\email[]{michal.silkowski@fuw.edu.pl}
\affiliation{Faculty of Physics, University of Warsaw, Pasteura 5, 02-093 Warsaw, Poland} 
\author{Jacek Komasa}
\affiliation{Faculty of Chemistry, Adam Mickiewicz University, Uniwersytetu Pozna{\'n}skiego 8, 61-614 Pozna{\'n}, Poland}
\author{Mariusz Puchalski}
\affiliation{Faculty of Chemistry, Adam Mickiewicz University, Uniwersytetu Pozna{\'n}skiego 8, 61-614 Pozna{\'n}, Poland}
\author{Krzysztof Pachucki}
\affiliation{Faculty of Physics, University of Warsaw, Pasteura 5, 02-093 Warsaw, Poland}

\begin{abstract}
We perform highly accurate calculation of the leading order QED correction to the ground electronic state of molecular hydrogen. Numerical results are obtained for a grid of the internuclear distances $R=0-10$ au with the relative precision of about 10\textsuperscript{-8}. The major numerical uncertainty of previous QED results [K.~Piszczatowski \emph{et al.},~JCTC~\textbf{5},~3039 (2009)] has been eliminated. Nevertheless, the discrepancy with measurements in~HD at~the level of 1.9 $\sigma$ persists.
\end{abstract}
\maketitle
 
\section{Introduction}

Quantum electrodynamic effects (QED) in atomic and molecular spectra are very difficult to determine computationally,
despite the fact that the exact formulas are well known~\cite{Bethe1957}. For this reason, they are often only roughly estimated based on hydrogenic results \cite{Eides:01}.
However, to obtain transition energies for few-electron systems with accuracy comparable to modern spectroscopic measurements, a high-precision computational method that accounts 
for a complete leading-order QED has to be employed.
So far, such calculations have been performed only for atoms with up to four electrons
\cite{Drake1992,Drake2011,Schwartz1961,Korobov1999,Korobov2012a,Korobov2019,Pachucki2010,LiYan,LiKP,BeKP}, and only for the simplest molecule, i.e., two-electron molecular hydrogen (H$_2$ and its deuterated and tritiated isotopologues) \cite{h22009}.   

Recent measurements of several rovibrational P- and R-branch transitions in the fundamental and overtone bands of the HD
molecule have reached unprecedented sub-MHz uncertainty \cite{Fast2020,Amsterdam2018,Amsterdam2019,Diouf:20,Amsterdam2022,Hefei2020,Hefei2022,Campargue2022,Caserta2021,Grenoble2022}.
It exceeds by at least an order of magnitude the accuracy of theoretical predictions including relativistic and QED contributions \cite{Ea,Ena,h2narel,h2rel,h2spectre}.
Moreover, systematic discrepancies of 1.4-1.9$\sigma$ are observed between calculated and experimental values, the origin of which is currently not clear. 
To improve theoretical predictions, at first, it is necessary to determine more precisely the leading $m \alpha^5$ QED
correction \cite{h22009}. Its numerical uncertainty is comparable to the estimate of unknown finite nuclear mass
(nonadiabatic) QED effects. Therefore, the development of a computational method that significantly reduces such numerical
inaccuracies is an indispensable step towards advancing the present theory of the hydrogen molecule to a higher level of accuracy.

In this work we perform high-precision calculations of leading $m \alpha^5$ QED correction in the ground electronic
state of a hydrogen molecule, using the Born-Oppenheimer (BO) approximation, thus omitting the nonadiabatic effect.
Our goal is to improve the accuracy of previous results \cite{h22009} by at least two orders of magnitude, including
the most computationally demanding Bethe logarithm term. To accomplish this, we employ explicitly correlated Gaussian
(ECG) basis functions and perform extensive variational optimization over all nonlinear parameters. The decisive advantage
of the ECG method is that the underlying integrations are manageable and very fast in numerical evaluation due to the
compact formulas for matrix elements of the nonrelativistic Hamiltonian, and they involve only well-known error function and elementary ones. Even though Gaussian functions
have the drawback of improper short-range form and fail to correctly describe the Kato cusp, it can be overcome with
a sufficiently large and well-optimized ECG basis set together with dedicated regularization techniques that accelerate the
convergence of singular operators. Furthermore, additional ECG integrals that arise as a consequence of the regularization
can be efficiently evaluated by means of dedicated numerical quadrature. Increasing the accuracy of the Bethe logarithm requires also
the development of efficient optimization algorithms, employing larger bases and a denser grid in the Schwartz integral
method~\cite{Schwartz1961} evaluated for a wide range of internuclear distances. In addition, it is important to derive
leading asymptotic terms, which are crucial for fitting the contours in numerical integration. All this considerable
effort is vital in laying the foundation for the future determination of non-adiabatic QED effects, which are the bottleneck
limiting the current accuracy of theoretical predictions for rovibrational energy levels of hydrogen molecule isotopologues.

\section{Leading order QED correction}

In this work we assume the adiabatic approximation (clamped nuclei), so that the total molecular
wavefunction $\Psi$ is a product of the electronic and nuclear parts,
\begin{equation}
	\Psi = \chi(\vec R)\,\phi(\vec r_1, \vec r_2; R).
\end{equation}
The leading $m\alpha^5$ QED correction 
\begin{align}
	E^{(5,0)} &= \langle \chi \lvert {\cal{E}}^{(5,0)}(R) \rvert \chi \rangle
\end{align}
to the molecular level is obtained by averaging the potential $\mathcal{E}^{(5,0)}(R)$ of Eq.~(\ref{E50}) 
with the nuclear wavefunction $\chi$; see \cite{h2spectre} for details. The QED potential
${\cal{E}}^{(5,0)}$ for a two-electron diatomic molecule can be compactly represented as~\cite{Bethe1957},
\begin{align}
	\label{E50}
	{\cal{E}}^{(5,0)}(R)&= \frac{4}{3}\Big( \frac{19}{30} - 2 \ln \alpha - \ln k_0 \Big) \sum_{i,X} Z_X \langle \delta^3(\vec r_{iX}) \rangle \nonumber \\
	& + \Big( \frac{164}{15} + \frac{14}{3} \ln \alpha \Big) \langle \delta^3(\vec r_{12}) \rangle
	- \frac{14}{3} \frac{1}{4\pi} \Big\langle \frac{1}{r_{12}^{3}} \Big\rangle_{\varepsilon}
\end{align}
where $R$ is the internuclear distance, $Z_X$ is the charge of nucleus $X$, the expectation value $\langle\ldots\rangle$
stands for integration over electronic degrees of freedom with the nonrelativistic wave function $\phi$,
$\ln k_0$ is the Bethe logarithm~\cite{Bethe1947} (see Eq. (\ref{lnk0}) below), and the last term is the Araki-Sucher
correction~\cite{Araki1957,Sucher1958} with $\langle r_{ij}^{-3} \rangle_{\varepsilon}$ denoting the following limit:
\begin{align}
	\bigg\langle \frac{1}{r_{ij}^{3}} \bigg\rangle_{\varepsilon} &\equiv 
\lim_{\varepsilon\rightarrow0}\left[\left\langle\frac{\Theta(r_{ij}-\varepsilon)}{r_{ij}^3}\right\rangle +
(\gamma_E+\ln \varepsilon)\big\langle 4 \pi \delta( \vec r_{ij})\big\rangle\right].\label{AS}
\end{align}
The symbol $\gamma_E$ denotes the Euler-Mascheroni constant, and $\Theta(x)$ 
is the Heaviside step function.

\section{Bethe logarithm}
At the level of Born-Oppenheimer approximation, Bethe logarithm enters as the $R$-dependent electronic quantity, defined as the following ratio of matrix elements~\cite{Bethe1947},
\begin{equation}
\label{lnk0}
\ln k_0 \equiv \frac{\langle \vec \nabla (\mathcal{H} - \mathcal{E}) \ln (2(\mathcal{H} - \mathcal{E})) \vec \nabla \rangle}{\langle \vec \nabla (\mathcal{H} - \mathcal{E}) \vec \nabla \rangle},
\end{equation}
with the electronic Hamiltonian,
\begin{equation}
	\mathcal{H} \equiv -\frac{1}{2} \sum_{i} \nabla^2_i -\sum_{i,X} \frac{Z_X}{r_{iX}} + \frac{1}{r_{12}} + \frac{Z_A Z_B}{R},
\end{equation}
and $\mathcal{E}$ its lowest energy eigenvalue,
\begin{equation}
	\mathcal{H} \phi(\vec r_1, \vec r_2; R) = \mathcal{E}(R) \phi(\vec r_1, \vec r_2; R).
\end{equation}
It can be represented in terms of the integral~\cite{LiKP,BeKP},
\begin{equation}
\label{lnk0int}
\ln k_0 = \frac{1}{\cal D}\,\int_0^1 dt\, \frac{f(t)-f_0 - f_2\,t^2}{t^3}  
\end{equation}
with
\begin{equation}
f(t) = -\biggl\langle\vec \nabla\,\frac{k}{k+\mathcal{H} - \mathcal{E}}\,\vec \nabla\biggr\rangle, \qquad t = \frac{1}{\sqrt{1+2\,k}}.
\label{ft}
\end{equation}
In the BO approximation, the current operator is purely electronic,
\begin{align}
	\vec \nabla &= \vec \nabla_1 + \vec \nabla_2\,, 
\intertext{and the denominator}
	{\cal D} &= 2\,\pi \sum_{i,X} \langle \delta^3(\vec r_{iX}) \rangle,
\end{align}
where the index $i$ runs over electrons and $X$ over nuclei.
The function $f(t)$ in Eq.~(\ref{lnk0int}) has the following expansion around $t=0$,
\begin{equation}
	\label{ftexp}
	f(t) = f_0 + f_2\; t^2 + f_3\; t^3 + (f^{l}_{4} \ln t + f_{4})\; t^4  + \mathcal{O}(t^5)\,,
\end{equation}
with the coefficients
\begin{eqnarray}
	\label{coeffs}
	f_0 &=& -\langle \vec \nabla^2 \rangle, \quad f_2 = -2\,{\cal D}\,, \quad	f_3 = 8\,{\cal D}, \quad f^{l}_{4} = 16\,{\cal D}, \nonumber \\
	f_{4} &=& 4 \Bigg[ \sum_{i,X} \Big\langle \frac{1}{r^4_{iX}} \Big\rangle_{\varepsilon} + \sum_{\substack{ (i,X),(j,Y) \nonumber \\
(i,X)\ne(j,Y)}} \Big\langle \frac{\vec{r}_{iX} \vec{r}_{jY}}{r_{iX}^3 r_{jY}^3} \Big\rangle \Bigg] \nonumber \\
 &-& 2\,{\cal D}\Big(1-4(1+\ln 4)\Big),
\end{eqnarray}
where
\begin{align}
	\label{rm4}
	\bigg\langle \frac{1}{r_{ij}^{4}}\bigg\rangle_{\varepsilon} &\equiv \lim_{\varepsilon\rightarrow0}\Bigg[ \Bigg\langle\frac{\Theta(r_{ij}-\varepsilon)}{r_{ij}^4}\Bigg\rangle - \frac{\langle 4\pi \delta(\vec r_{ij})\rangle}{\varepsilon} \\ \nonumber
											& +2\bigg\langle 4\pi\delta(\vec r_{ij})\frac{\partial }{\partial r_{ij}} \bigg\rangle (\gamma_E+\ln \varepsilon)\Bigg].
\end{align}

In the early days of quantum electrodynamics the calculation of the Bethe logarithm for systems beyond hydrogen-like
atoms, even with a few percent accuracy, emerged as a challenging task \cite{Pekeris1957,Salpeter1957,Salpeter1961}.
In the approach presented by Schwartz \cite{Schwartz1959a,Schwartz1959b,Schwartz1961} the evaluation of $f(t)$ was
reformulated into the second-order problem of finding $\vec \phi_1$ satisfying the following inhomogeneous differential equation:
\begin{align}
	\label{fphi1}
	(\mathcal{E} - \mathcal{H}-k)\vec \phi_1 = \vec \nabla \phi,
\end{align}
so that the sum over states $f(t)$ is simply given by $\vec \phi_1$ as
\begin{align}
	\label{fkphi1}
	f(t) &= k \langle \vec \phi_1 \lvert \vec \nabla \rvert \phi \rangle.
\end{align}
This is equivalent to finding a stationary value of the following Ritz functional
\begin{align}
	w(k) &= 2\langle \phi \lvert \vec \nabla \rvert \vec \phi_1 \rangle + \langle \vec \phi_1 \lvert \mathcal{E} - \mathcal{H}-k \rvert \vec \phi_1 \rangle,
\end{align}
which with the stationarity condition
\begin{align}
	\delta w(k) &\equiv 2\langle \phi \lvert \vec \nabla \rvert \delta \vec \phi_1 \rangle + 2 \langle \vec \phi_1 \lvert \mathcal{E} - \mathcal{H}-k \rvert \delta \vec \phi_1 \rangle = 0
\end{align}
recovers (\ref{fphi1}) due to the arbitrariness of variation $\delta \vec \phi_1$.
Such a formulation allows for variational computation of $w(k)$. For the sake of evaluating the integral (\ref{lnk0int}), the matrix element (\ref{fkphi1}) has to be minimized on a grid of values of $k$, bearing in mind the manifest dependence of $\vec \phi_1$ on $k$.

\section{Method}

For the purpose of variational calculations of the resolvent, we follow with the decomposition of $f(t)$, into $f_{\parallel}(t)$ and $f_{\perp}(t)$, which emerges from the decomposition of $\vec \nabla$,
\begin{eqnarray}
 \vec \nabla &=& \vec n\,(\vec n \cdot \vec \nabla) - \vec n \times (\vec n \times \vec \nabla).
\end{eqnarray}
Namely,
\begin{eqnarray}
	f(t) &=& f_{\parallel}(t) + f_{\perp}(t), \\
	f_{\parallel}(t) &=& -\biggl\langle Q_{\parallel} \,\frac{k}{k+\mathcal{H} - \mathcal{E}}\, Q_{\parallel} \biggr\rangle, \nonumber \\
	f_{\perp}(t) &=& -\biggl\langle \vec Q_{\perp} \,\frac{k}{k+\mathcal{H} - \mathcal{E}}\, \vec Q_{\perp} \biggr\rangle,
\end{eqnarray}
where
\begin{eqnarray}
Q_{\parallel} &=&  \vec n\,\cdot \vec \nabla,\\
\vec Q_{\perp} &=& \vec n \,\times (\vec n \times \vec \nabla).
\end{eqnarray}
Because the ground state is of $\Sigma_g^{+}$ symmetry, this entails dipole connected intermediate states of $\Sigma^{+}_u$ and $\Pi^{-}_u$ symmetry in the resolvents in $f_{\parallel}$ and $f_{\perp}$, respectively.

The external wavefunction $\phi$ is expanded in ECG basis functions; see Eq.~(\ref{Psiexpansion}) below.
For a series of basis sizes $N\,\in\,\{64,128,256,512,768,1024\}$ and for 55 values of internuclear distance
$R$ in the range $0.05 - 10$\,au, we have minimized $f_{\parallel}(t)$ and $f_{\perp}(t)$ on a uniform grid ($\Delta t = 0.01$).
We followed the heuristic approach of Refs.~\cite{naqediso,h2naqed} and set the size of intermediate state basis as
$N'=2N$ for $t\le0.1$ and $N'=\frac{3}{2}N$ otherwise.

To efficiently evaluate the integral (\ref{lnk0int}) we split the integration domain into two regions: $t\in\left<0,t_{\mathrm{crit}}\right>$ and $t\in(t_{\mathrm{crit}},1\rangle$. 
The high-$t$ region is free of singularities, and $f(t)$ can be efficiently integrated numerically. For this purpose we
have optimized $f(t>t_{\mathrm{crit}})$ on a uniform $t$-grid with the spacing of 0.01.
Because $f(1)$ satisfies the generalized Thomas-Reiche-Kuhn (TRK) sum rule \cite{Yan2006},
\begin{equation}
	\langle \vec \nabla (\mathcal{H} - \mathcal{E})^{-1} \vec \nabla \rangle = -3,
\end{equation}
we have utilized it in practical computations to deduce the completeness of the intermediate state basis and estimate the numerical uncertainty of $f(t)$.
Then, the integral is evaluated by means of interpolating the numerical data with a high-order polynomial (typically of
order 18). With the largest bases considered, the uncertainty resulting from numerical integration in the high-$t$ region
is of order $10^{-11}$, which is a few orders of magnitude less than the uncertainty coming from the
integration of the low-$t$ region, the latter being of critical importance for high accuracy of the final value of $\ln k_0$.

Primarily, in low-$t$ region a strong numerical cancellation between $f(t)$ and leading asymptotic terms $f_0$ and $f_2 t^2$ occurs. We
emphasize that for high photon momenta~($t~\rightarrow~0$) the integrand in the integral definition of $\ln k_0$, see Eq. ($\ref{lnk0int}$), is dominated by $\sim t^{-3}$, so that two leading terms of Taylor expansion have to be subtracted from $f(t)$ to render it integrable.
For that reason, $f(t<t_{\mathrm{crit}})$ cannot be evaluated with high-accuracy in the same way as in high-$t$ region,
and we resort to elementary, term-by-term, analytical integration of the asymptotic expansion ($\ref{ftexp}$).
Nonetheless, to achieve a highly precise final value of $\ln k_0$, inclusion only of known terms (up to
$\sim t^4$) is insufficient and higher-order coefficients of low-$t$ asymptotics of $f(t)$ have to be added. They are determined by fitting them from the numerical
data from the range $(t_{\mathrm{cut}},t_{\mathrm{crit}})$ as,
\begin{equation}
	\delta f(t) \equiv f(t) - \big( f_0 + f_2\; t^2 + f_3\; t^3 + (f^{l}_{4} \ln t + f_{4})\; t^4 \big),
\end{equation}
with the functional form of fitted expansion deduced from the known behavior of $f(t)$ for the hydrogen atom~\cite{Pachucki1993,Gavrila1970},
\begin{equation}
	\frac{\delta f(t)}{t^5} = \sum_{m=0}^M t^m ( f_m + f^{l}_{m} \ln t ),
\end{equation}
with fixed $f^{l}_{m}=0$ for $m$ even.
For very low values of $t$ a $t_{\mathrm{cut}}$ cutoff discards numerical points
$f(t<t_{\mathrm{cut}})$, which are of insufficient numerical accuracy, due to the presence of $t^{-3}$ acting as a
weighting factor greatly enhancing the demand on the numerical accuracy of $f(t)$ as $t \rightarrow 0$.
Resultantly, values of $f(t<t_{\mathrm{cut}})$ have to be discarded completely and cannot be used even for the purpose of fitting higher order terms of low-$t$ asymptotics.

Ultimately, $t_{\mathrm{cut}}$, $t_{\mathrm{crit}}$, and $M$ are adjustable parameters, which are tuned with the purpose of reaching the final value of $\ln k_0$, such that it presents weak sensitivity to their change. Fluctuations of $\ln k_0$ due the change of those parameters around their optimal values serve as an uncertainty estimation.
Typically, optimal values lie in the range $t_{\mathrm{cut}} \in (0.02, 0.06)$, $t_{\mathrm{crit}}\in(0.12, 0.23)$, and $M\in(2, 6)$, with pronounced tendency of preffered larger $t_{\mathrm{crit}}$ whenever higher fit order $M$ is demanded.
\subsection{ECG method}
In our calculations we utilize an explicitly correlated Gaussian (ECG) basis,
\begin{align}
	\phi &= \sum_i c_i \phi_i(\vec r_1, \vec r_2), \nonumber \label{Psiexpansion}\\
	\phi_i &= (1 \pm \mathcal{P}_{A \leftrightarrow B})(1 \pm \mathcal{P}_{1 \leftrightarrow 2}) \nonumber \\
		 &\times e^{-a_{12} r^2_{12} - a_{1A} r^2_{1A} - a_{1B} r^2_{1B} - a_{2A} r^2_{2A} - a_{2B} r^2_{2B}}.
\end{align}
Direct inclusion of the interelectronic distance $r_{12}$ in the exponent of trial wavefunction renders it a
two-particle,
two-center geminal, \emph{explicitly correlated} basis.
The primary virtue of the ECG basis is that all the requisite integrals for calculations of nonrelativistic energy and $f(t)$ can be evaluated very efficiently.
All required matrix elements can be expressed in terms of linear combinations of the following ECG integrals:
\begin{align}
	\label{intI}
	&I(n_1,n_2,n_3,n_4,n_5)\equiv\int\frac{d^{\mathrm{3}}r_1}{\pi^{3/2}} \int\frac{d^{\mathrm{3}}r_2}{\pi^{3/2}} r_{1A}^{n_1} r_{1B}^{n_2} r_{2A}^{n_3} r_{2B}^{n_4} r_{12}^{n_5} \\ \nonumber
	&\times e^{ -a_{1A} r^2_{1A} -a_{1B} r^2_{1B} -a_{2A} r^2_{2A} -a_{2B} r^2_{2B} -a_{12} r^2_{12}},
\end{align}
with integer $n_i$ and real parameters $a$.
It is clear that differentiation of this integral with respect to the given nonlinear parameter $a$ raises the appropriate index $n_i$ by 2.
Consequently, disjoint families of ECG integrals arise.
The first family is termed \emph{regular} ECG integrals and is defined by $\Omega_1=0,2,4,\ldots$ and non-negative
even integers $n_i$, such that $\sum_i n_i \le \Omega_1$.
Among these integrals, the following master integral plays a pivotal role:
\begin{eqnarray}
	I(0,0,0,0,0) &=&X^{-3/2}\,e^{-R^2\,Y/X},
\end{eqnarray}
where
\begin{eqnarray}
	X&=&(a_{1A}+a_{1B}+a_{12})(a_{2A}+a_{2B}+a_{12})-a_{12}^2,\\
	Y&=&(a_{1A}+a_{1B})\,a_{2A}\,a_{2B} + (a_{2A}+a_{2B})\,a_{1A}\,a_{1B} \nonumber \\
	 &+& a_{12}(a_{1A}+a_{2A})(a_{1B}+a_{2B}).
\end{eqnarray}
All the other \emph{regular} ECG integrals can be generated by differentiation of $I(0,0,0,0,0)$ over $a$-parameters.

Another family, \emph{Coulomb} ECG integrals, permits a single odd index $n_i \ge -1$, with $\sum_i n_i \le \Omega_2$
($\Omega_2 = -1,1,3,\ldots$), and analogously to \emph{regular} ECG integrals, all \emph{Coulomb} ECG integrals can be generated by differentiation
over $a$-parameters of appropriate master integrals,
\begin{align}
	I(-1_i)&=\frac{1}{X\sqrt{X_i}}\,e^{-R^2\,Y/X}\,F\Bigg[R^2\bigg(\frac{Y_i}{X_i}-\frac{Y}{X}\bigg)\Bigg],
\end{align}
where $I(-1_i)$ denotes $I$ with $n_i=-1$ and other indices equal zero, $X_i\equiv\partial_{a_i}X,~Y_i\equiv\partial_{a_i}Y$, and $F(x)\equiv\mathrm{erf}(x)/x$.

In contrast to atomic ECG integrals, the molecular ones have no known analytic form whenever two or more indices are odd.
Nevertheless, such \emph{extended} ECG integrals arise either as a consequence of the regularization of expectation
values, as described in the next Subsection, or as matrix elements of the coefficients of high-momentum asymptotic expansion of $f(t)$.
Fortunately, such \emph{extended} integrals can be efficiently evaluated by means of numerical quadrature.
When there is no logarithm in the integrand, the quadrature relies on the following Gaussian integral transform,
\begin{align}
	\frac{1}{r^n} &= \frac{2}{\Gamma(n/2)} \int_0^{\infty} dt\,t^{n-1}\,e^{-r^2 t^2}, ~n>0.
\end{align}
In this case, the integral (\ref{intI}) can be represented as
\begin{align}
	& I(n_1-n,n_2,n_3,n_4,n_5) = \nonumber \\
	& \frac{2}{\Gamma(n/2)} \int_0^{\infty} \mathrm{d}y\,y^{n-1}\,I(n_1,n_2,n_3,n_4,n_5) \bigg|_{a_{1A}\rightarrow a_{1A}+y^2}.
\end{align}
With the help of the variable transformation~$y=-1+1/x$, which reduces the integration domain of \emph{extended} ECG
integral to a finite interval of $(0,1)$, the integral can be readily evaluated by the generalized extended Gaussian quadrature with logarithmic end-point singularities \cite{gausext},
\begin{align}
	\label{quad}
	& \int_0^1 dx \big[ W_1(x) + \ln(x) W_2(x) \big] = \nonumber \\
      & \sum_i^m w_i \big[ W_1(x_i) + \ln (x_i) W_2(x_i) \big];
\end{align}
thus,
\begin{align}
	& I(n_1-n,n_2,n_3,n_4,n_5) = \nonumber \\
	& \frac{2}{\Gamma(n/2)} \sum_{i=1}^m w_i \frac{(1-x_i)^{n-1}}{x_i^{n+1}}\,I(n_1,n_2,n_3,n_4,n_5)\bigg|_{a_{1A}\rightarrow a_{1A}+y_i^2}.
\end{align}
In the case of integrals involving logarithms, we utilize the following transforms:
\begin{align}
	\label{lnrrm1}
	\frac{\ln r}{r} &= -\frac{1}{\sqrt{\pi}}\,\int_0^{\infty} dt\,(2\ln t + \gamma_E + \ln 4)\,e^{-r^2 t^2}, \\
	\frac{\ln r}{r^2} &= -\int_0^{\infty} dt\,t\,(2\ln t + \gamma_E)\,e^{-r^2 t^2}.
\end{align}

This approach can be straightforwardly generalized to double quadrature over Coulomb ECG integral over two different nonlinear parameters, which
allows us to calculate integrals with three odd indices. Such integrals arise during calculations of large photon momentum asymptotic coefficients of $f(t)$.

\subsection{Regularization}

According to (\ref{E50}), the $\mathcal{E}^{(5,0)}$ correction appears as a deceivingly simple sum of expectation values.
Those expectation values, however, are of operators that are rather nontrivial.
They probe the wavefunction in the vicinity of Coulombic singularities, as in the case of Araki-Sucher
correction or even exactly pointwise at those singularities in the case of Dirac delta functions.

It is well-known that the ECG basis cannot reproduce the correct asymptotic behavior of the wavefunction around
electron-electron and electron-nucleus coalescence points (cusp conditions) and resultantly yields slow convergence of
expectation values of singular operators with a very local integral kernel.
This disadvantage of the ECG basis can be circumvented by utilizing strong operator identities, which probe the wavefunction more
globally, making expectation values much less sensitive to the local deficiencies of the wavefunction \cite{Drachman1981,Pachucki2005,naqed,naqediso}.
Here we introduce three such identities, the first two of which are vital for the high-accuracy of
the QED potential because enter $\mathcal{E}^{(5,0)}$ directly whereas the last one enables accurate
evaluation of the asymptotic coefficient $f_4$,
\begin{eqnarray}
	\langle 4 \pi \delta^3(\vec r_{ab}) \rangle &=& 2 \mu_{ab} \bigg[ 2 V^{(1)}_{ab} - R^{(1)}_{ab} \bigg], \nonumber \\
	\label{rm3}
	\bigg\langle \frac{1}{r^3_{ab}} \bigg\rangle_{\varepsilon} &=& \left(1 + \gamma_E \right) \langle 4 \pi \delta^3(\vec r_{ab}) \rangle + 2 \mu_{ab} \bigg[ 2 \tilde{V}^{(1)}_{ab} - \tilde{R}^{(1)}_{ab} \bigg], \nonumber \\
	\bigg\langle \frac{1}{r^4_{ab}} \bigg\rangle_{\varepsilon} &=& \mu_{ab} \bigg[ -2{V}^{(2)}_{ab} + {R}^{(2)}_{ab} \pm \langle 12 \pi \delta^3(\vec r_{ab}) \rangle \bigg].
\end{eqnarray}
In the above, $\mu_{ab}=\frac{m_a m_b}{m_a + m_b}$ is the reduced mass of pair of particles $a$ and $b$, and in the last formula '$+$'
should be taken for particles with the same, and '$-$' with opposite charges, respectively. Furthermore,
\begin{eqnarray}
	V_{ab}^{(n)} &\equiv& \left\langle \frac{1}{r_{ab}^n} (E - V) \right\rangle, \\
	\tilde{V}_{ab}^{(n)} &\equiv& \left\langle \frac{\ln r_{ab}}{r_{ab}^n} (E - V) \right\rangle, \\
	R_{ab}^{(n)} &\equiv& - \sum_{i=1,2} \left\langle \vec \nabla_i \frac{1}{r_{ab}^n} \vec \nabla_i \right\rangle, \\
	\tilde{R}_{ab}^{(n)} &\equiv& - \sum_{i=1,2} \left\langle \vec \nabla_i \frac{\ln r_{ab}}{r_{ab}^n} \vec \nabla_i \right\rangle,
\end{eqnarray}
where $r_{ab}$ pertains to either electron-electron or electron-nucleus coordinates.
This regularization procedure is pivotal for achieving well-converged, high-accuracy expectation values of singular operators with ECG functions \cite{h2rel,h2naqed}.

\section{Results}

\begin{table*}[ht]
	\centering
	\caption{Convergence of the electronic BO energy (${\cal{E}}$) and terms contributing to $\mathcal{E}^{(5,0)}$
	with increasing basis size $N$ at $R=1.4$\,au.
		For fixed $N$, the uncertainty of $\ln k_0$ originates from uncertainties of the $f_{i>4}$. Here, as well as for all other values of $R$, this uncertainty dominates over the uncertainty resulting from extrapolation to the complete basis set limit. All presented digits of Dirac delta expectation values and $\langle r^{-3}_{12} \rangle_{\varepsilon}$ are accurate.
}
	\begin{tabular}{d{4.0} d{3.20} d{3.12} d{3.12} d{3.12} d{3.12} }
	\hline
	\hline
	
	\multicolumn{1}{c}{$N$} & \multicolumn{1}{c}{${\cal{E}}$} & \multicolumn{1}{c}{$\sum_{i,X} \langle \delta^3(\vec r_{iX}) \rangle$} & \multicolumn{1}{c}{$\langle \delta^3(\vec r_{12}) \rangle$} & \multicolumn{1}{c}{ $\langle r^{-3}_{12} \rangle_{\varepsilon}$} & \multicolumn{1}{c}{$\ln k_0$} \Tstrut \\
	\hline \Bstrut \\
	\input{Re.tex} \\
	\hline
	\hline
	\end{tabular}
	\label{tab:Re}
\end{table*}

The hydrogen molecule in its ground state dissociates into H$(1s)$+H$(1s)$.
Therefore, in our calculations we benefit from the fact that the analytical form of $f(t)$, the essential part of the integrand
of the integral representation of $\ln k_0$, is known exactly for the hydrogen-like atom~\cite{Pachucki1993,Gavrila1970}:
\begin{align}
	f^{\mathrm{H}}(t) \equiv -384\frac{t^5}{(1+t)^8(2-t)} {}_2F_1(4,2-t,3-t;\xi),
\end{align}
where ${}_2F_1$ is the hypergeometric function in standard notation \cite{Abramowitz1988} and 
\begin{align}
	\xi=[(1-t)/(1+t)]^2, \qquad	t = Z/\sqrt{-2(\mathcal{E}-k)}.
\end{align}
Resultantly, the numerical value of the Bethe logarithm for the ground state of the hydrogen atom ($Z=1$, $\mathcal{E}=-1/2$) is
known with many-digit accuracy \cite{Drake1990},
\begin{align}
	\label{lnkh}
	\ln k_0(\mathrm{H})  =&\ 2.984\,128\,555\,765\,498\,\ldots
\end{align}
The dominating contribution to the Bethe logarithm comes from the high momenta of photon excitation~\cite{Bethe1950}, thus involving highly excited continuum states.
Therefore, it is very insensitive to the details of perturbation of electronic structure as induced by the presence of another hydrogen atom.
As a result, we expect that not only $f(t)$ but also individual terms of its Taylor expansions should be relatively close to those of
$f^{\mathrm{H}}(t)$ for all but very small values of $R$.

The greater accuracy of $\ln k_0$ near the equilibrium is a consequence of purposeful computational focus on optimization of $f(t)$
by employing larger basis sets ($N$=768,1024). Moreover, $\ln k_0(R)$ changes rather slowly for
$R>5$ au, and this region is much less important in view of averaging $\mathcal{E}^{(5,0)}$ with nuclear wavefunction;
therefore, the largest size of external basis used there was only $N=512$.
Deterioration of $\ln k_0$ uncertainty as $R \rightarrow 0$ is the consequence of large uncertainty of fitted expansion
in the low-$t$ region, due to its high sensitivity to the fitting parameters.

\begin{figure}[ht]
\includegraphics[width=\columnwidth]{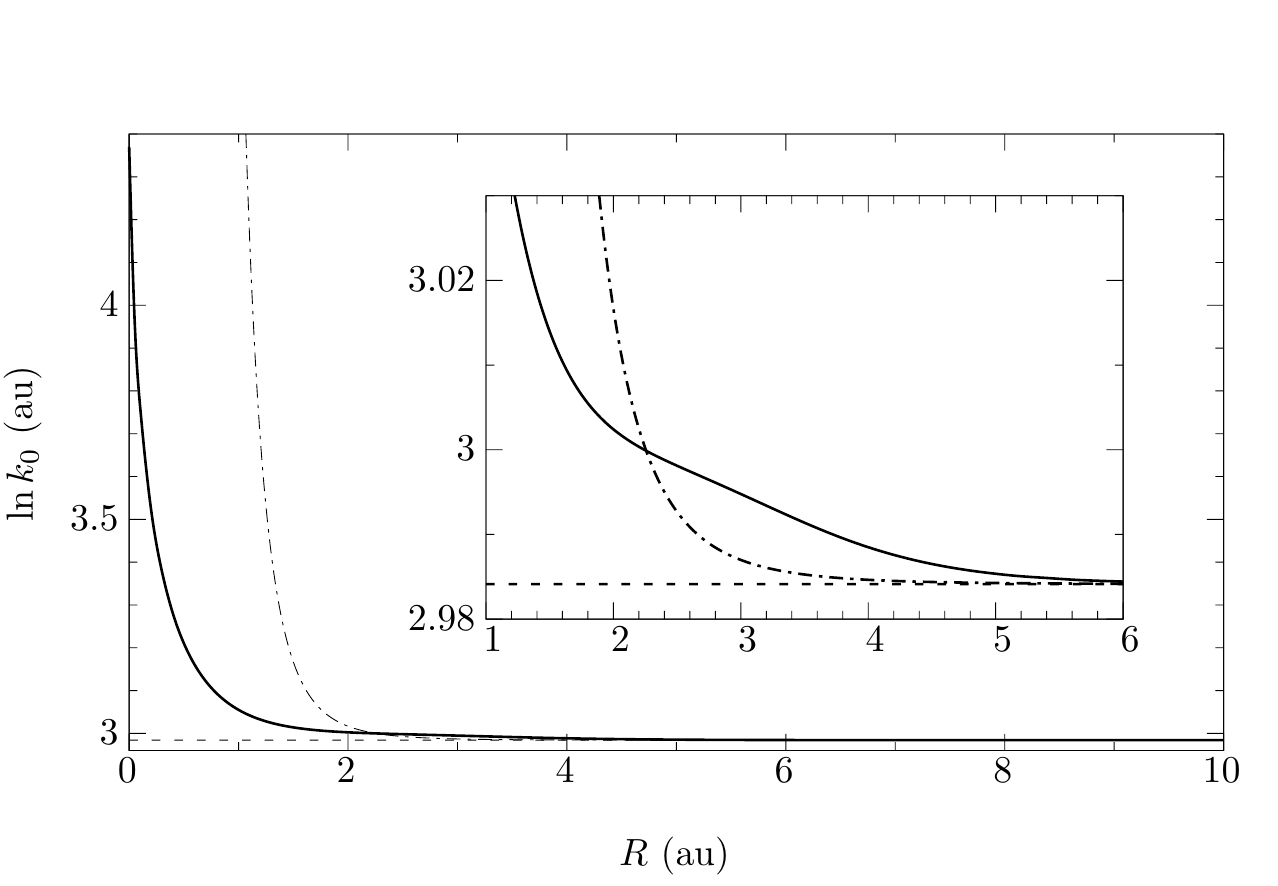}
\caption{The solid black line represents $\ln k_0(R)$, Eq. (\ref{lnk0int}). A very rapid drop from the united-atom (helium) limit can
be observed. The inset displays nontrivial dependence of $\ln k_0$ as a function of $R$ in the region $R=1-6$ au, which
prevents the use of a simple exponential decay fit. 
The dash-dotted line presents the only known leading order long-range asymptotics,
Eq.~(\ref{lnkasymp}). The dashed line corresponds to the asymptotic hydrogenic limit.}
\label{fig:lnk}
\end{figure}

Although $\ln k_0$ as a function of $R$ changes rapidly from its united-atom helium value to the hydrogenic one in
a manner resembling exponential decay, it exhibits nontrivial behavior in the region around equilibrium internuclear distance, see Fig. $\ref{fig:lnk}$.
Consequently, commonly used one-parameter approximation to the $R$-behavior of the Bethe logarithm,
\begin{equation}
	\ln k_0(\mathrm{H}) + \big[\ln k_0(\mathrm{He})~-~\ln k_0(\mathrm{H})\big] e^{-a R},
\end{equation}
when both united-atom and dissociation limits are usually known much more accurately (as is the
case with H$_2$), is far from sufficient for high-precision theoretical predictions.

\begin{figure}[ht]
\includegraphics[width=\columnwidth]{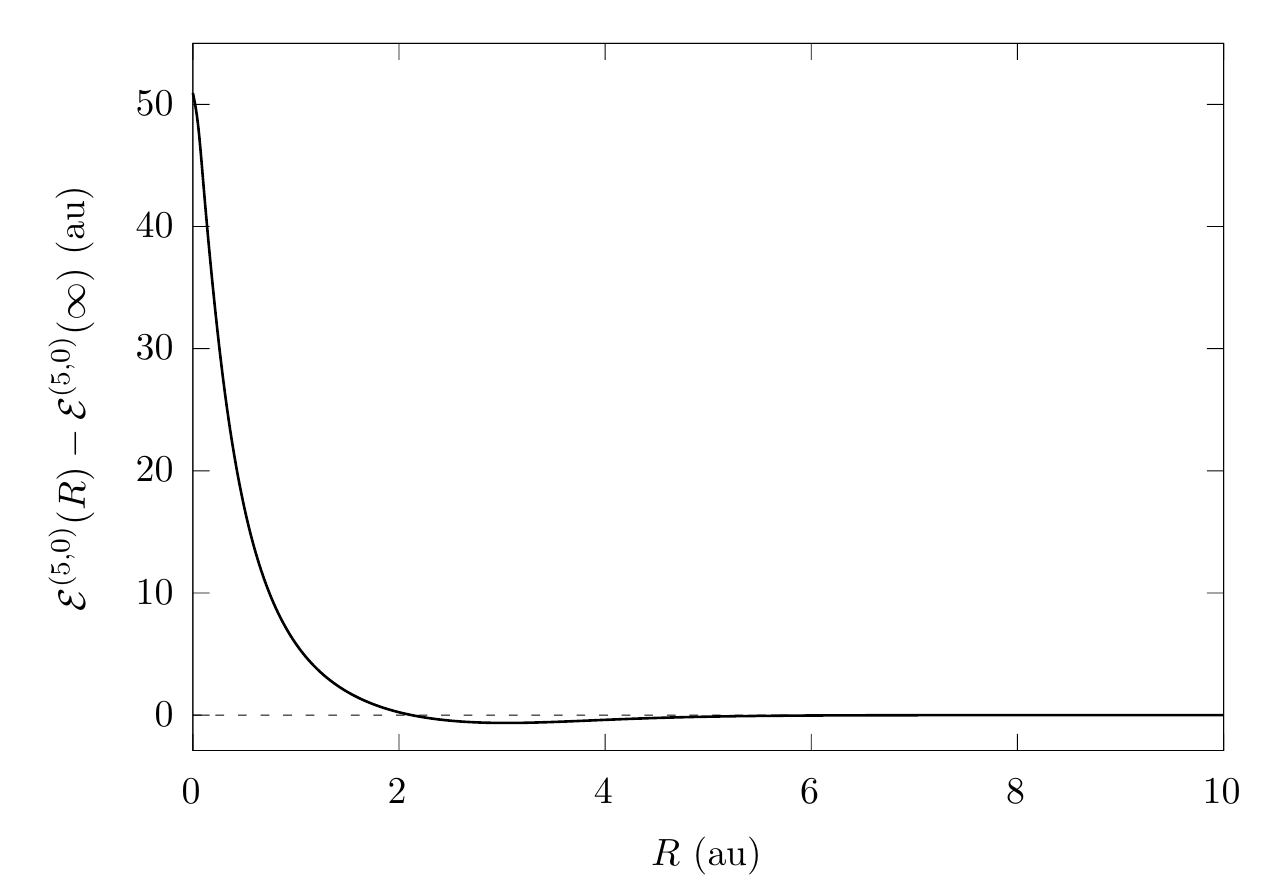}
\caption{The solid black line represents $\mathcal{E}^{(5,0)}$ of Eq. (\ref{E50}) as a function of $R$ with the
asymptotic hydrogenic limit subtracted.}
\label{fig:E50}
\end{figure}

Convergence with the basis size and comparison to the literature of Dirac delta, Araki-Sucher and Bethe logarithm at $R=1.4$ au is presented in Table
$\ref{tab:Re}$. Final values of ${\cal{E}}^{(5,0)}$ and its essential components are presented in Table $\ref{tab:final}$, whereas its
$R$-behavior is plotted in Fig. $\ref{fig:E50}$.

Ultimately, we recognize the obtained absolute Bethe logarithm accuracy of $3 \times 10^{-8}$ with $N \sim 1000$ to be
satisfactory, especially in view of its proximity to the absolute accuracy of the Araki-Sucher term which is of similar magnitude (about $10^{-8}$).

\subsection{Long-range asymptotics of the Araki-Sucher correction}
At the dissociation limit only the first term of Eq. (\ref{E50}) persists, so that
\begin{align}
	{\cal{E}}^{(5,0)}(\infty) &= \frac{4}{3}\Big( \frac{19}{30} - 2 \ln \alpha - \ln k_0 (\mathrm{H}) \Big) \frac{2}{\pi} \\
					  &= 6.357\,448\,103\,05(2),
\end{align}
with $1/\alpha = 137.035\,999\,206(11)$~\cite{CODATA2018} and the value of $\ln k_0 (\mathrm{H})$ given by Eq.  (\ref{lnkh}).

We have found that $\langle r_{12}^{-3} \rangle_{\varepsilon}$, as evaluated according to Eq. (\ref{rm3}) using a single
quadrature utilizing transform Eq. (\ref{lnrrm1}), exhibits very clear $\sim m^{-6}$ convergence, with
$m$ being the number of quadrature nodes, Eq. (\ref{quad}). Therefore, a very accurate Richardson extrapolation is possible, which allows for
effortless improvement of the accuracy by roughly 2 orders of magnitude. Although the accuracy of those operators is usually
good enough even with $m=40$, we have found such an acceleration of convergence very useful and necessary for the
sake of as accurate as possible comparison to analytical long-range asymptotic expansion of $\langle r_{12}^{-3}
\rangle_{\varepsilon}$, which reads \cite{Lach},
\begin{align}
	\label{pasymp}
	\Big\langle \frac{1}{r^{3}_{12}} \Big\rangle_{\varepsilon} &= \frac{1}{R^3} + \frac{6}{R^5} + \frac{75}{R^7} - C_6\,\frac{10}{R^8} +
	\frac{1575}{R^9} + \mathcal{O}(R^{-10}),
\end{align}
where $C_6 = 6.499026705405\ldots$ is the leading order coefficient (dipole-dipole) of the long-range asymptotics of dispersion energy.

Due to the high accuracy of our data, we have attempted to fit higher-order terms by subtracting the asymptotics up to order
$R^{-9}$ and fitting a series in powers of $1/R$.
We have found such fits to be sensitive to both the expansion order and the number of points used.
We noticed that the leading coefficient of the fit is oscillating around a value given by (\ref{pasymp}) as the expansion order is incremented by one. This observation
strongly suggests the presence of higher order terms with large coefficients, and we estimate the next term to be $-3400/R^{10}$ with 50\% uncertainty.
A meaningful comparison with even higher-order terms would require data at $R>20$~au or higher accuracy (better than $10^{-12}$) of our large-$R$ results, which entails costly
optimization of even larger basis sets and is of little practical significance.

\subsection{Long-range asymptotics of the Bethe logarithm}
The Bethe logarithm in H$_2$ is known to behave asymptotically as
\begin{align}
	\label{lnkasymp}
	\ln k_0(R) = \ln k_0(\mathrm{H}) + \frac{L_6}{R^6} + \mathcal{O}(R^{-8}) \quad \mathrm{as} \quad R \rightarrow \infty,
\end{align}
with $L_6=2.082\,773\,197$~\cite{h22009}.
Comparison with our data suggests that this asymptotic expansion is not sufficient to accurately describe the behavior
of $\ln k_0 (R)$ for $R$ as large as 6-10~au, in spite of its numerical value rapidly approaching that of the hydrogen atom.
In particular, this two-term asymptotic expansion diverges from numerical data by as much as 27, 157, and 568\% at $R=10,8$, and 6 au, respectively.
This suggests the significance of higher order terms. Large magnitude of their coefficients is tentatively confirmed by our fitting attempts.

\section{Conclusions}
We have performed highly accurate calculations of QED effects in the ground state of molecular hydrogen.
Due to the accuracy of order of $10^{-8}$, which is 2 to 3 orders better than previous calculations \cite{h22009}, major
numerical uncertainty of the QED effects on the molecular levels has been eliminated.
Nevertheless, the shift of about 0.03 MHz with respect to Ref.~\cite{h22009} is below the level of existing discrepancies with measured transition energies (1.4-1.9~$\sigma\approx 2$\,MHz) in the HD molecule.
Fully nonadiabatic QED calculations~\cite{h2naqed,naqediso} performed for the lowest levels of H$_2$ ($\nu=0$, $J=0$)
reduce the uncertainty of the QED contribution to the level of 5\,kHz. Together with the results obtained in present work, this indicates the significance of
nonadiabatic QED effects in hydrogen molecule and its isotopologues. These effects can be calculated with the help of nonadiabatic perturbation theory (NAPT) \cite{napt}, which is planned
in the near future, and present work can be regarded as a first step toward this goal.

The obtained results have already been included in the updated version (v7.4) of the publicly available computer code \textsc{H2spectre} \cite{h2spectre}.

\section*{Acknowledgements}
MP and MS acknowledge support from the National Science Center (Poland) under Grant No. 2019/34/E/ST4/00451.
MS acknowledges funding support from Grant No. 2020/36/T/ST2/00605 as well as by computing grant from PL-Grid Infrastructure.

\bibliography{h2qed}

\newpage

\begin{table*}[b]
	\centering
	\footnotesize
	\caption{Terms contributing to $\mathcal{E}^{(5,0)}$ as a function of $R$. Uncertainties originate from
	extrapolation to the complete basis set.}
	\begin{tabular}{d{3.4} d{3.16} d{3.16} d{3.16} d{3.16} d{3.16}}
	\hline
	\hline
	
	\multicolumn{1}{c}{$R$} & \multicolumn{1}{c}{$\sum_{i,X} \langle \delta^3(\vec r_{iX}) \rangle$} &
	\multicolumn{1}{c}{$\langle \delta^3(\vec r_{12}) \rangle$} & \multicolumn{1}{c}{$\langle r^{-3}_{12}
	\rangle_{\varepsilon}$} & \multicolumn{1}{c}{$\ln k_0$} & \multicolumn{1}{c}{${\cal{E}}^{(5,0)}(R)-{\cal{E}}^{(5,0)}(\infty)$} \Tstrut \\
	\hline \Bstrut \\
	\input{final.tex} \\
	\hline
	\hline
	\end{tabular}
	\label{tab:final}
\end{table*}

\end{document}

%% file: Re.tex
64                                            & -1.174\,474\,384\,972\,363       &  0.919\,300\,2223   & 0.016\,739\,9980    & 0.414\,497\,7238  &  3.018\,421(77)      \\ 
128                                           & -1.174\,475\,663\,522\,751       &  0.919\,333\,4195   & 0.016\,742\,9651    & 0.414\,364\,3945  &  3.018\,549\,0(37)   \\ 
256                                           & -1.174\,475\,712\,366\,731       &  0.919\,335\,9183   & 0.016\,743\,2525    & 0.414\,346\,7147  &  3.018\,561\,00(43)  \\ 
512                                           & -1.174\,475\,714\,135\,081       &  0.919\,336\,1813   & 0.016\,743\,2745    & 0.414\,345\,0950  &  3.018\,563\,264(40) \\ 
768                                           & -1.174\,475\,714\,210\,245       &  0.919\,336\,2021   & 0.016\,743\,2769    & 0.414\,344\,8871  &  3.018\,563\,400(20) \\ 
1024                                          & -1.174\,475\,714\,218\,617       &  0.919\,336\,2099   & 0.016\,743\,2776    & 0.414\,344\,8224  &  3.018\,563\,453(28) \\ 
\multicolumn{1}{c}{$\infty$}                  & -1.174\,475\,714\,221(1)         &  0.919\,336\,214(3) & 0.016\,743\,2780(4) & 0.414\,344\,79(3) &  3.018\,563\,480(38) \\
\multicolumn{1}{c}{Ref. \cite{h22009}}        & -                                &  0.919\,34(1)       & 0.016\,74(1)        & 0.414\,30(1)        &  3.018\,55(1)           \\
\multicolumn{1}{c}{Ref. \cite{Matyus2022}}    & -1.174\,475\,714(1)              &  -                  & -                   & -                   &  3.018\,55(3)               \\
\multicolumn{1}{c}{Ref. \cite{h2rel}\footnote{Evaluated with 1024-term rECG basis, ${\cal{E}}$ without extrapolation to complete basis set}}         & -1.174\,475\,714\,203            &  0.919\,336\,206(7) & 0.016\,743\,2783(5) & -                   &  -                   \\
\multicolumn{1}{c}{Ref. \cite{h2rel,h2bo}\footnote{Evaluated with James-Coolidge wavefunction}}    & -1.174\,475\,714\,220\,443\,4(5) &  0.919\,336\,211(2) & 0.016\,743\,2783(3) & -                   &  -                   \\

%% file: final.tex
0.0 	    &	7.24171727400(4)\footnote{\label{frolov}Ref. \cite{frolov}}      & 0.106345370635(1)\footref{frolov}    & 0.989273545024(1)\footref{frolov}   & 4.3701602230703(3)\footnote{Ref. \cite{Korobov2019}}  & 50.930716724136(3)  \\
0.05 	    &	6.4890437210(4)      &  0.105003821909(7)    &    0.9881318794(5)   & 3.9610(2)      &     48.362(2)  \\
0.1 	    &	5.770597278(3)       &  0.10157055016(10)    &    0.98432455(2)     & 3.7616(2)      &     43.700(2)  \\
0.2 	    &	4.568931914(3)       &  0.09136818987(9)     &    0.96590419(2)     & 3.52291(5)     &     34.5292(3)  \\
0.3 	    &	3.671598383(2)       &  0.0798419070(2)      &    0.93232441(2)     & 3.37749(2)     &     27.07582(5)  \\
0.4 	    &	3.007974280(3)       &  0.0688732774(1)      &    0.88704623(2)     & 3.279584(5)    &     21.33818(2)  \\
0.5 	    &	2.512087713(3)       &  0.0591374522(2)      &    0.83484011(4)     & 3.210230(2)    &     16.950265(6)  \\
0.6 	    &	2.135554242(4)       &  0.0507632372(2)      &    0.77971886(2)     & 3.1595728(9)   &     13.569055(3)  \\
0.7 	    &	1.844767256(4)       &  0.0436605313(2)      &    0.72449332(2)     & 3.1218242(4)   &     10.9319924(9)  \\
0.8 	    &	1.616559028(4)       &  0.0376678752(4)      &    0.67093263(4)     & 3.0933129(2)   &     8.8483668(4)  \\
0.9 	    &	1.434821980(8)       &  0.0326147204(4)      &    0.62004755(4)     & 3.07157925(5)  &     7.18120486(9)  \\
1.0 	    &	1.288195834(4)       &  0.0283452755(4)      &    0.57234019(4)     & 3.05490864(6)  &     5.8317574(1)  \\
1.1 	    &	1.168538455(4)       &  0.0247256344(4)      &    0.52798840(4)     & 3.04206959(5)  &     4.72813145(7)  \\
1.2 	    &	1.069918931(4)       &  0.0216439932(3)      &    0.48697212(3)     & 3.03215694(6)  &     3.81734866(9)  \\
1.3 	    &	0.987949457(3)       &  0.0190083084(3)      &    0.44915639(3)     & 3.02449314(6)  &     3.05986972(9)  \\
1.4 	    &	0.919336214(3)       &  0.0167432780(4)      &    0.41434479(3)     & 3.01856348(4)  &     2.42581252(4)  \\
1.4011    &	0.918645801(5)       &  0.0167201832(4)      &    0.41397774(4)     & 3.01850633(3)  &     2.41943366(4)  \\
1.5 	    &	0.861572961(6)       &  0.0147874126(5)      &    0.38231287(4)     & 3.01397237(6)  &     1.89232027(8)  \\
1.6 	    &	0.812728773(5)       &  0.0130904713(5)      &    0.35282881(5)     & 3.01041312(6)  &     1.44170996(7)  \\
1.7 	    &	0.771298712(4)       &  0.0116113186(4)      &    0.32566573(3)     & 3.00764631(6)  &     1.06015438(6)  \\
1.8 	    &	0.736097174(4)       &  0.0103161797(3)      &    0.30060873(3)     & 3.00548408(5)  &     0.73673165(5)  \\
1.9 	    &	0.706180622(4)       &  0.0091772346(2)      &    0.27745861(2)     & 3.00377871(6)  &     0.46273095(6)  \\
2.0 	    &	0.680790767(4)       &  0.0081714960(4)      &    0.25603365(4)     & 3.00241335(9)  &     0.23113917(9)  \\
2.1 	    &	0.659312208(4)       &  0.0072799102(4)      &    0.23617006(3)     & 3.00129678(5)  &     0.03625438(5)  \\
2.2 	    &	0.641240392(4)       &  0.0064866435(2)      &    0.21772165(3)     & 3.0003573(2)   &     -0.1266062(1)  \\
2.3 	    &	0.626157038(4)       &  0.0057785088(3)      &    0.20055918(3)     & 2.9995393(3)   &     -0.2613324(2)  \\
2.4 	    &	0.613711019(4)       &  0.0051445018(2)      &    0.18456954(2)     & 2.99880014(8)  &     -0.37119725(7)  \\
2.5 	    &	0.603603308(5)       &  0.0045754299(3)      &    0.16965457(3)     & 2.9981078(3)   &     -0.4589970(2)  \\
2.6 	    &	0.595575025(6)       &  0.0040636043(3)      &    0.15573008(2)     & 2.9974395(2)   &     -0.5271620(2)  \\
2.7 	    &	0.589397878(6)       &  0.0036025887(9)      &    0.14272450(8)     & 2.99677927(9)  &     -0.57784529(8)  \\
2.8 	    &	0.584866587(6)       &  0.0031869850(9)      &    0.13057766(9)     & 2.99611714(9)  &     -0.61299348(7)  \\
2.9 	    &	0.581792978(6)       &  0.0028122509(3)      &    0.11923957(3)     & 2.9954475(3)   &     -0.6344010(2)  \\
3.0 	    &	0.580001468(5)       &  0.0024745485(2)      &    0.10866836(2)     & 2.9947706(3)   &     -0.6437534(2)  \\
3.2 	    &	0.579608561(5)       &  0.0018975786(7)      &    0.08969278(6)     & 2.9933998(2)   &     -0.6326256(1)  \\
3.4 	    &	0.582448355(5)       &  0.0014346013(5)      &    0.07342502(5)     & 2.9920403(2)   &     -0.5916362(1)  \\
3.6 	    &	0.587399916(3)       &  0.0010683465(2)      &    0.05967099(2)     & 2.9907405(2)   &     -0.5317100(2)  \\
3.8 	    &	0.593491441(3)       &  0.0007837408(3)      &    0.04822793(3)     & 2.9895459(4)   &     -0.4623141(4)  \\
4.0 	    &	0.599942218(4)       &  0.0005669019(3)      &    0.03886333(2)     & 2.9884879(1)   &     -0.3910095(7)  \\
4.2 	    &	0.606187475(4)       &  0.0004049339(2)      &    0.03131461(2)     & 2.9875819(9)   &     -0.3231954(8)  \\
4.4 	    &	0.611872454(2)       &  0.0002861635(2)      &    0.025304362(9)    & 2.9868296(3)   &     -0.2621756(8)  \\
4.6 	    &	0.616819342(2)       &  0.00020046794(8)     &    0.020560798(7)    & 2.9862180(4)   &     -0.2094973(3)  \\
4.8 	    &	0.620980474(2)       &  0.00013946740(5)     &    0.016835178(4)    & 2.9857306(2)   &     -0.1654339(2)  \\
5.0 	    &	0.624391499(3)       &  0.00009651477(4)     &    0.013912356(3)    & 2.9853497(3)   &     -0.1294587(3)  \\
5.2 	    &	0.627132992(2)       &  0.00006652495(4)     &    0.011614267(3)    & 2.9850545(5)   &     -0.1006249(4)  \\
5.4 	    &	0.629303303(2)       &  0.00004571971(2)     &    0.009798327(2)    & 2.9848280(5)   &     -0.0778397(4)  \\
5.6 	    &	0.631001614(2)       &  0.00003135473(3)     &    0.008353048(2)    & 2.9846557(3)   &     -0.0600270(3)  \\
5.8 	    &	0.632318799(2)       &  0.00002147077(1)     &    0.007192666(1)    & 2.9845254(3)   &     -0.0462145(2)  \\
6.0 	    &	0.633333462(2)       &  0.00001468696(3)     &    0.006251868(2)    & 2.9844270(3)   &     -0.0355683(3)  \\
6.5 	    &	0.634946115(2)       &  0.000005668122(8)    &    0.0045649148(5)   & 2.9842755(4)   &     -0.0186014(4)  \\
7.0 	    &	0.635760799(2)       &  0.000002183104(3)    &    0.0034749552(2)   & 2.9842020(5)   &     -0.0099569(4)  \\
7.5 	    &	0.636169868(2)       &  0.000000840032(2)    &    0.00272930901(9)  & 2.9841664(4)   &     -0.0055487(5)  \\
8.0 	    &	0.636376357(2)       &  0.0000003230160(9)   &    0.00219493423(5)  & 2.9841490(4)   &     -0.0032671(6)  \\
8.5 	    &	0.636482278(2)       &  0.0000001241243(5)   &    0.00179784702(2)  & 2.9841401(4)   &     -0.0020520(6)  \\
9.0 	    &	0.636538096(2)       &  0.0000000476605(5)   &    0.001494408972(5) & 2.9841350(3)   &     -0.0013766(3)  \\
9.5 	    &	0.636568630(2)       &  0.0000000182846(2)   &    0.001257399493(6) & 2.9841331(4)   &     -0.0009818(6)  \\
10.0 	    &	0.636586108(2)       &  0.0000000070084(2)   &    0.001068988784(2) & 2.9841314(4)   &     -0.0007357(6)  \\
11.0 	    &	0.636603270(2)       &  0.00000000102659(2)  &    0.000792906742(2) & & \\
12.0 	    &	0.636610686(2)       &  0.00000000014980(1)  &    0.000605092990(2) & & \\
13.0 	    &	0.636614385(2)       &  0.00000000002177(1)  &    0.000472597263(2) & & \\
14.0 	    &	0.636616412(2)       &  0.00000000000315(1)  &    0.000376332776(2) & & \\
15.0 	    &	0.636617594(2)       &  0.00000000000045(2)  &    0.000304652082(2) & & \\
16.0 	    &	0.636618315(2)       &  0.00000000000006(1)  &    0.000250149428(2) & & \\
17.0 	    &	0.636618771(2)       &  0.00000000000001(1)  &    0.000207953638(2) & & \\
18.0 	    &	0.636619068(2)       &  0.00000000000000(1)  &    0.000174767149(2) & & \\
19.0 	    &	0.636619267(2)       &  0.00000000000000(1)  &    0.000148301511(2) & & \\
20.0 	    &	0.636619402(2)       &  0.00000000000000(1)  &    0.000126933698(2) & & \\
\multicolumn{1}{c}{$\infty$}  &	0.63661977236 &  0.0                  &    0.0 & 2.984128555765 & 0.0 \\